# Generalizing Cross Products and Maxwell's Equations to Universal Extra Dimensions

A.W. McDavid[a][*] and C.D. McMullen[b][†]

[a] *Department of Physics and Astronomy, Louisiana State University,
Baton Rouge, LA 70803, USA*

[b] *Department of Physics, Louisiana School for Math, Science, and the Arts,
Natchitoches, LA 71457, USA*

## Abstract

In the past five years, there has been considerable research on the collider phenomenology of TeV-scale extra compact dimensions which are universal – i.e. accessible to all the Standard Model fields. We consider in detail a fundamental conceptual issue in the higher-dimensional theory: Since the cross product is only physically meaningful in three spatial dimensions, how do traditional vector formulas involving the cross product or curl operation generalize in the case of universal extra dimensions? For example, is angular momentum physically meaningful, and, if so, how is it related to the position and momentum vectors? We find that if angular velocity, torque, angular momentum, and magnetic field, for example, are regarded as anti-symmetric second-rank tensors instead of pseudovectors, we can construct tensor forms of the standard cross product and curl equations that do generalize to higher dimensions. Although Maxwell's equations have been generalized to extra dimensions in terms of differential forms, we show that our simple technique casts the generalized Maxwell's equations in a form very similar to the standard vector calculus. We also provide direct integrals for the fields in a practical tensor calculus form. We explore the electric and magnetic fields in detail to show that there are no fundamental conceptual problems hidden in the theory – e.g. how the Lorentz force can be meaningfully expressed in terms of the velocity and magnetic field when the traditional cross product and right-hand rule do not apply. Finally, we explore the effects that compactification has on the effective form of Maxwell's equations, which could be useful for direct tests of Coulomb's law.

---

[*] email: amcdav3@lsu.edu
[†] email: cmcmullen@lsmsa.edu

## 1. Introduction

Since 1998, when Arkani-Hamed, Dimopoulos, and Dvali motivated extra compact dimensions much larger than the inverse Planck scale as a novel solution to the hierarchy problem [1], there has been much research on the astrophysical/cosmological [2] and collider phenomenology [3] of sub-millimeter size string-inspired [4] extra compact dimensions accessible only to gravitons. In 2001, when Appelquist, Cheng, and Dobrescu proposed that one or more superstring-inspired extra compact dimensions may be universal [5] – i.e. accessible to all Standard Model (SM) fields – if the compactification scale is at least 300 GeV, there followed further research on the collider phenomenology [6] of $TeV^{-1}$-size universal extra compact dimensions. The formalism for computing the effects of the universal extra dimensions is as follows. First, the $(N+1)$-D spacetime Lagrangian density $L_{N+1}$ is expressed via straightforward generalization of the usual $(3+1)$-D Standard Model Lagrangian density $L_{SM}$. The $(N+1)$-D fields are then Fourier expanded about the compact dimensions, and the effective $(3+1)$-D Lagrangian density $L_{eff}$ is obtained by integrating over the compact coordinates. This yields the Feynman rules in the effective $(3+1)$-D theory. One feature of models with universal extra dimensions is the presence of Kaluza-Klein (KK) excitations of the SM fields in the effective $(3+1)$-D theory resulting from the Fourier expansions. These KK excitations have masses on the order of the compactification scale $1/R$, which must be at least ~350-400 GeV in order for the KK excitations to have escaped detection at current high-energy colliders.

In this straightforward generalization of the SM Lagrangian density to higher dimensions, Maxwell's inhomogeneous equations are immediately obtained in concise tensor form in terms of the higher-dimensional field strength tensor **F** [7]-[8]. However, it is well-known that the cross product is only valid in $(3+1)$-D and $(7+1)$-D, but physically meaningful only in $(3+1)$-D [9]. Correspondingly, the traditional right-hand rule associated with the cross product seems ambiguous in higher dimensions – e.g. any vector in the $x_3 x_4$ plane is orthogonal to both $\hat{\mathbf{x}}_1$ and $\hat{\mathbf{x}}_2$. However, the $(3+1)$-D Maxwell's equations, expressed in terms of the electric and magnetic fields, involve the curl operations $\vec{\nabla} \times \vec{\mathbf{E}}$ and $\vec{\nabla} \times \vec{\mathbf{B}}$, and the force experienced by an electric charge in the presence of a magnetic field involves the cross product $\vec{\mathbf{F}}_M = q\vec{\mathbf{v}} \times \vec{\mathbf{B}}$. How can the common practice of generalizing the SM Lagrangian density be reconciled with the obvious problems associated with generalizing the cross product to higher dimensions?

We assert that in all the traditional $(3+1)$-D curl operations and cross products in physics, such as the definition of torque $\vec{\tau} = \vec{\mathbf{r}} \times \vec{\mathbf{F}}$, one of the vectors involved is actually an anti-symmetric second rank tensor. This common identity problem is a consequence of the coincidence that, in $(3+1)$-D, vectors and anti-symmetric second rank tensors each have three independent components. If angular velocity, torque, angular momentum, and magnetic field, for example, are regarded as anti-symmetric second-rank tensors instead of pseudovectors, the tensor forms of the standard cross product and curl equations do generalize to higher dimensions.

Magnetic field is naturally interpreted as an anti-symmetric second-rank tensor $\underline{\mathbf{B}}$ [8] in that it is the spatial subset of the field strength tensor $\underline{\mathbf{F}}$, while electric field $\vec{\mathbf{E}}$ is naturally a vector formed from the temporal components of $\underline{\mathbf{F}}$. While Maxwell's equations have been generalized to extra dimensions in terms of differential forms [10], we show how our simple tensor forms of the cross product and curl operation express the generalized Maxwell's equations similar to the standard vector calculus form. We explore in detail the differences in the nature of the fields – i.e. the vector electric field and tensor magnetic field $\underline{\mathbf{B}}$ – to show that there no underlying conceptual problems in the higher-dimensional theory associated with the inapplicability of the traditional cross product and right-hand rule.

For example, Gauss's law normally involves the scalar products $\vec{\mathbf{E}} \cdot d\vec{\mathbf{A}}$ and $\vec{\mathbf{B}} \cdot d\vec{\mathbf{A}}$ and Stokes's theorem normally involves the scalar products $\vec{\mathbf{E}} \cdot d\vec{\mathbf{s}}$ and $\vec{\mathbf{B}} \cdot d\vec{\mathbf{s}}$. Although Gauss's law and Stokes's theorem have previously been generalized to extra dimensions in terms of differential forms, we note that when applied to Maxwell's equations these laws must be expressed differently for the vector electric field $\vec{\mathbf{E}}$ and tensor magnetic field $\underline{\mathbf{B}}$. In $(N+1)$-D, the differential area element $d\vec{\mathbf{A}}$ is promoted to a higher-dimensional differential element with $N-1$ spatial dimensions, and the line element $d\vec{\mathbf{s}}$ is similarly promoted to $N-2$ spatial dimensions. In general, the higher-dimensional analog of the differential path length can be expressed as an anti-symmetric second-rank tensor, while the higher-dimensional analog of the surface area can be expressed as a vector. This means that the higher-dimensional analog of $\vec{\mathbf{B}} \cdot d\vec{\mathbf{A}}$ involves a tensor magnetic field $\underline{\mathbf{B}}$ and a vector differential element. Similarly, $\vec{\mathbf{E}} \cdot d\vec{\mathbf{s}}$ involves a vector electric field $\vec{\mathbf{E}}$ and a tensor differential element. While $\vec{\mathbf{E}} \cdot d\vec{\mathbf{A}}$ remains a scalar product between vectors, $\underline{\mathbf{B}} \cdot d\underline{\mathbf{s}}$ becomes a contraction of two tensors.

Although electrodynamics must macroscopically appear effectively $(3+1)$-D after compactification, we feel that it is important to analyze Maxwell's equations in detail in $(N+1)$-D in order to show that the usual prescription for generalizing the $(3+1)$-D Lagrangian density to higher dimensions is fundamentally sound. The usual duality between the electric and magnetic fields is lost in higher dimensions, where the magnetic field $\underline{\mathbf{B}}$ has $N(N-3)/2$ more independent components than the electric field $\vec{\mathbf{E}}$ has. As a result, the introduction of magnetic monopoles to complete the usual symmetry between the fields is not as aesthetically appealing as it is in $(3+1)$-D. We also provide direct integrals for the fields, which are more practical than the usual geometric extensions of Gauss's law and Stokes's theorem when there is not much symmetry.

Our paper is organized as follows. In Sec. 2, we generalize the traditional cross products and curl operations in physics to higher dimensions by identifying appropriate vectors as anti-symmetric second-rank tensors. In Sec. 3, we discuss the form that the differential line element $d\vec{\mathbf{s}}$, area element $d\vec{\mathbf{A}}$, and volume element $dV$ assume when Gauss's law and Stokes's theorem are generalized to higher dimensions. In Sec. 4, we express Maxwell's equations in higher dimensions in differential and integral form in terms of the tensor magnetic field $\underline{\mathbf{B}}$ and vector electric field $\vec{\mathbf{E}}$. In Sec. 5, we discuss force laws, including Coulomb's law and the Biot-Savart law, in higher dimensions. We discuss compactification on an $S^1/Z_2$ orbifold in Sec. 6, and in Sec. 7 we illustrate how this compactification causes deviations from the usual inverse-square dependence in Coulomb's law, which could be useful for direct tests of Coulomb's law. We draw our conclusions in Sec. 8.

**2. Cross Products**

The cross product originates as a mathematical construction to represent physical phenomena in which two quantities with three independent components, such as the position $\vec{r}$ and momentum $\vec{p}$ of a classical point-particle, combine to form a third quantity with three independent components, as in the case of angular momentum $\vec{L} = \vec{r} \times \vec{p}$. These three-component quantities are intuitively interpreted to be vectors. In this traditional sense, the cross product of two vectors $\vec{A}$ and $\vec{B}$ yields a pseudovector $\vec{C}$ with components given by

$$C_k = \varepsilon_{ijk} A_i B_j \tag{1}$$

However, in $(3+1)$-D the pseudovector $\vec{C}$ could just as well be an anti-symmetric second-rank tensor $\underline{C}$ with components given by

$$C_{ij} = A_i B_j - A_j B_i \tag{2}$$

Observe that the components of the anti-symmetric second-rank tensor $\underline{C}$ are related to the components of the pseudovector $\vec{C}$ via

$$\begin{aligned} C_{ij} &= \varepsilon_{ijk} C_k \\ C_k &= \frac{1}{2!} \varepsilon_{ijk} C_{ij} \end{aligned} \tag{3}$$

Alternatively, if $\vec{B}$ is a pseudovector while $\vec{A}$ and $\vec{C}$ are vectors, where $\vec{C} = \vec{A} \times \vec{B}$, then the pseudovector $\vec{B}$ may be expressed as an anti-symmetric second-rank tensor $\underline{B}$ with components given by

$$C_j = A_i B_{ji} \tag{4}$$

In this case, the components of the anti-symmetric second-rank tensor $\underline{B}$ and the pseudovector $\vec{B}$ are again related by Eq. (3). Notice that $A_i B_{ij}$ in the tensor formulation corresponds to the $j^{\text{th}}$ component of the cross product $\vec{B} \times \vec{A}$. Since the cross product and tensor formulations are equivalent in $(3+1)$-D, preference is traditionally given to working with a pseudovector rather than an anti-symmetric second-rank tensor. This leads to the convenient right-hand rule for relating the directions of the vectors.

However, as previously stated, Eq. 1 does not generalize to higher dimensions. If the SM fields propagate into one or more universal extra dimensions, then it is necessary to find a means of generalizing formulas involving the cross product and curl for physical quantities, such as angular momentum and magnetic force, in order for the higher-dimensional theory to be physically meaningful at the fundamental level. Since the usual cross product is ruled out in this case, we turn to the alternative constructions involving one anti-symmetric second rank tensor and two vectors. Ultimately, any attempt to construct a physically viable cross product that yields one vector in terms of two others in higher dimensions breaks down due to a lack of information – two $N$-component vectors do not give rise to a uniquely specified third $N$-component vector satisfying the desirable vector algebra in $N > 3$ spatial dimensions, as illustrated conceptually by the ambiguity associated with any attempt to apply the traditional right-hand rule. For example, in $(3+1)$-D only a vector parallel to $\pm\hat{\mathbf{x}}_3$ is orthogonal to both $\hat{\mathbf{x}}_1$ and $\hat{\mathbf{x}}_2$, but in $(4+1)$-D any vector in the $x_3 x_4$ plane is orthogonal to both $\hat{\mathbf{x}}_1$ and $\hat{\mathbf{x}}_2$.

In contrast to Eq. 1, Eq.'s 2 and 4 relate two $N$-component vectors via one $N \times N$ anti-symmetric second-rank tensor with $N(N-1)/2$ independent components. The tensor constructions, Eq.'s 2 and 4, do generalize to higher dimensions. As the dimensionality of space grows from $N$ to $N+1$, vectors gain one component while anti-symmetric second-rank tensors gain $N$ components – the same amount of additional information as an extra $N$-component vector. It is the additional components of the tensor form that make the tensor construction work while the traditional cross product between vectors does not. In addition to matching components mathematically, there is a physical motivation for the tensor construction. For example, the magnetic field is more naturally interpreted as an anti-symmetric second-rank tensor than a vector in that it is the spatial subset of the field strength tensor $\underline{\mathbf{F}}$. We assert that all the traditional cross products and curls in physics actually involve one anti-symmetric second rank tensor and two vectors.

The following examples illustrate how standard cross products and curl operations in physics are expressed in terms of appropriate anti-symmetric second-rank tensors:

$$\begin{aligned}
\vec{\mathbf{L}} = \vec{\mathbf{r}} \times \vec{\mathbf{p}} &\rightarrow L_{ij} = x_i p_j - x_j p_i \\
\vec{\mathbf{v}} = \vec{\boldsymbol{\omega}} \times \vec{\mathbf{r}} &\rightarrow \upsilon_i = x_j \omega_{ji} \\
\vec{\mathbf{F}} = q\vec{\mathbf{E}} + q\vec{\mathbf{v}} \times \vec{\mathbf{B}} &\rightarrow F_i = qE_i + q\upsilon_j B_{ij} \\
\vec{\mathbf{B}} = \vec{\nabla} \times \vec{\mathbf{A}} &\rightarrow B_{ij} = \partial_i A_j - \partial_j A_i \\
\vec{\mathbf{S}} = \frac{1}{\mu_0} \vec{\mathbf{E}} \times \vec{\mathbf{B}} &\rightarrow S_i = \frac{E_j B_{ij}}{\mu_0}
\end{aligned} \quad (5)$$

The tensor angular momentum $\underline{\mathbf{L}}$ is particularly interesting since, even in the standard formulation, moment of inertia $\underline{\mathbf{I}}$ is a symmetric second-rank tensor with six independent components. The components of these tensors are related by

$$L_{ij} = \frac{1}{2!} \varepsilon_{ijk} \varepsilon_{\ell mn} I_{kn} \omega_{\ell m} \tag{6}$$

The tensor forms of the traditional cross products are derived via application of Eq. 3 along with the following useful relations:

$$\begin{aligned} \varepsilon_{ijk}\varepsilon_{ij\ell} &= 2!\delta_{k\ell} \\ \varepsilon_{ijk}\varepsilon_{\ell mk} &= \delta_{i\ell}\delta_{jm} - \delta_{im}\delta_{j\ell} \end{aligned} \tag{7}$$

Here, we show that the quantities $\underline{\mathbf{C}}$ in Eq. 2 and $\underline{\mathbf{B}}$ in Eq. 4 are indeed second-rank tensors based on their behavior under a coordinate transformation $\{x_i\} \to \{x'_i\}$:

$$\begin{aligned} C'_{ij} &= A'_i B'_j - A'_j B'_i & C'_j &= A'_i B'_{ji} \\ C'_{ij} &= \lambda_{ik}\lambda_{j\ell}(A_k B_\ell - A_\ell B_k) & \lambda_{jk} C_k &= \lambda_{i\ell} A_\ell B'_{ji} \\ C'_{ij} &= \lambda_{ik}\lambda_{j\ell} C_{k\ell} & \lambda_{jm}\lambda_{jk} C_k &= \lambda_{jm}\lambda_{i\ell} A_\ell B'_{ji} \\ & & C_m &= \lambda_{jm}\lambda_{i\ell} A_\ell B'_{ji} \\ & & B_{m\ell} &= \lambda_{jm}\lambda_{i\ell} B'_{ji} \end{aligned} \tag{8}$$

where the $\{\lambda_{ij}\}$ are the direction cosines. The usual vector algebra is preserved in the tensor formulation of the conventional cross product. For example, $C_{ij} C_{ij} = 2 A^2 B^2 \sin^2\theta$ provides the usual relation for the angle between $\vec{A}$ and $\vec{B}$. Also, the cross product of a vector with itself yields the null tensor: $\vec{A} \times \vec{A} \to A_i A_j - A_j A_i = 0$. The formula $\vec{A} \times (\vec{B} \times \vec{C}) = \vec{B}(\vec{A} \cdot \vec{C}) - \vec{C}(\vec{A} \cdot \vec{B})$ assumes different forms, depending on which pseudovectors are replaced with anti-symmetric second-rank tensors. For example, if $\vec{C}$ is a pseudovector while $\vec{A}$ and $\vec{B}$ are vectors, then in terms of the anti-symmetric second-rank tensor $\underline{\mathbf{C}}$, the triple product $\vec{A} \times (\vec{B} \times \vec{C})$ becomes an anti-symmetric second-rank tensor with components $B_k(A_i C_{jk} - A_j C_{ik})$.

The tensor form of the conventional cross product formulas naturally generalize to higher dimensions. For example, $L_{ij} = r_i p_j - r_j p_i$ becomes $L_{IJ} = r_I p_J - r_J p_I$, where the spatial indices $\{i,j\} \in \{1,2,3\}$ in $(3+1)$-D simply grow to spatial indices $\{I,J\} \in \{1,2,3,4\}$ in $(4+1)$-D.

Observe that Eq. 3 holds only in $(3+1)$-D. That is, in $(N+1)$-D, where $N > 3$, the components of the anti-symmetric second-rank tensor can not be related to the components of a vector. Namely, there is a mismatch of components: $N(N-1)/2$ for the anti-symmetric second rank tensor compared to $N$ for the vector. Also, while Eq. 3 does not hold in $(4+1)$-D, there is an analogous equation for expressing an anti-symmetric (henceforth, we describe a higher-rank tensor as antisymmetric if it is antisymmetric under the interchange of any two indices) third-rank tensor as a four-component vector: [‡]

$$\begin{aligned} C_{IJK} &= \varepsilon_{IJKL} C_L \\ C_L &= \frac{1}{3!} \varepsilon_{IJKL} C_{IJK} \end{aligned} \tag{9}$$

While it is very straightforward to generalize the conventional cross product equations, with the form of Eq. 5, by simply growing the indices, other equations involving these anti-symmetric second-rank tensors require additional care. For example, in $(4+1)$-D the equation relating angular momentum to moment of inertia becomes

$$L_{IJ} = \frac{\varepsilon_{IJKP} \varepsilon_{LMNP} I_{KN} \omega_{LM}}{2!} \tag{10}$$

When generalizing to $(4+1)$-D, it is useful to note that the $(4+1)$-D analog of Eq. 7 is

$$\begin{aligned} \varepsilon_{IJKL} \varepsilon_{IJKM} &= 3! \delta_{LM} \\ \varepsilon_{IJKL} \varepsilon_{IJMN} &= 2!(\delta_{KM} \delta_{LN} - \delta_{KN} \delta_{LM}) \\ \varepsilon_{IJKL} \varepsilon_{IMNP} &= \delta_{JM} \delta_{KN} \delta_{LP} - \delta_{JM} \delta_{KP} \delta_{LN} + \delta_{JN} \delta_{KP} \delta_{LM} \\ &\quad - \delta_{JN} \delta_{KM} \delta_{LP} + \delta_{JP} \delta_{KM} \delta_{LN} - \delta_{JP} \delta_{KN} \delta_{LM} \end{aligned} \tag{11}$$

In $(5+1)$-D, it is possible to express an anti-symmetric fourth-rank tensor in terms of a 5-component vector and, similarly, an anti-symmetric third-rank tensor in terms of an anti-symmetric second rank tensor with analogs of Eq. 9:

$$\begin{aligned} C_{I_5 J_5 K_5 L_5} &= \varepsilon_{I_5 J_5 K_5 L_5 M_5} C_{M_5} & C_{I_5 J_5 K_5} &= \frac{1}{2!} \varepsilon_{I_5 J_5 K_5 L_5 M_5} C_{L_5 M_5} \\ C_{M_5} &= \frac{1}{4!} \varepsilon_{I_5 J_5 K_5 L_5 M_5} C_{I_5 J_5 K_5 L_5} & C_{L_5 M_5} &= \frac{1}{3!} \varepsilon_{I_5 J_5 K_5 L_5 M_5} C_{I_5 J_5 K_5} \end{aligned} \tag{12}$$

where $\{I_5, J_5\} \in \{1,2,3,4,5\}$. The $(5+1)$-D analog of Eq. 11 is

$$\begin{aligned} \varepsilon_{I_5 J_5 K_5 L_5 M_5} \varepsilon_{I_5 J_5 K_5 L_5 N_5} &= 4! \delta_{M_5 N_5} \\ \varepsilon_{I_5 J_5 K_5 L_5 M_5} \varepsilon_{I_5 J_5 K_5 N_5 P_5} &= 3!(\delta_{L_5 N_5} \delta_{M_5 P_5} - \delta_{L_5 P_5} \delta_{M_5 N_5}) \\ \varepsilon_{I_5 J_5 K_5 L_5 M_5} \varepsilon_{I_5 J_5 N_5 P_5 Q_5} &= 2!(\delta_{K_5 N_5} \delta_{L_5 P_5} \delta_{M_5 Q_5} - \delta_{K_5 N_5} \delta_{L_5 Q_5} \delta_{M_5 P_5} + \delta_{K_5 P_5} \delta_{L_5 Q_5} \delta_{M_5 N_5} \\ &\quad - \delta_{K_5 P_5} \delta_{L_5 N_5} \delta_{M_5 Q_5} + \delta_{K_5 Q_5} \delta_{L_5 N_5} \delta_{M_5 P_5} - \delta_{K_5 Q_5} \delta_{L_5 P_5} \delta_{M_5 N_5}) \end{aligned} \tag{13}$$

---

[‡] Notice that the extended Levi-Civita symbol $\varepsilon_{IJKL}$ we are using here has four spatial indices (in contrast to the spacetime version $\varepsilon_{\mu\nu\rho\sigma}$, in which the contravariant and covariant forms differ in signs).

It is straightforward to generalize the usual vector algebra formulas to higher dimensions by first expressing the usual $(3+1)$-D equations in appropriate tensor form and applying Eq. 11, where applicable. For example, the angle between $\vec{A}$ and $\vec{B}$ can still be obtained from $C_{I_N J_N} C_{I_N J_N} = 2 A^2 B^2 \sin^2 \theta$, even in higher dimensions, where $\underline{C}$ is given by the higher-dimensional generalization of Eq. 2.

## 3. Differential Elements

In this section, we provide a tensor construction of the differential elements that closely parallels vector calculus, which will serve to distinguish between the vector and tensor nature of the electric and magnetic fields. This will also allow us to generalize Maxwell's equations and derive direct integrals for the fields in a form closely parallel to the usual vector calculus approach – an alternative to using differential forms, which have previously been used to generalize Gauss's law and Stokes's theorem to higher dimensions (which is conceptually powerful when there is ample symmetry).

We begin by discussing the ordinary differential elements in the usual three spatial dimensions. The line element $d\vec{s}$ is naturally expressed as a differential displacement vector with rectangular components $ds_i = dx_i$. The scalar form, i.e. the differential arc length $ds$, is then easily constructed via contraction: $ds = \sqrt{dx_i dx_i}$. The line element can also be expressed as an anti-symmetric second rank tensor $d\underline{s}$ with components $ds_{ij} = \varepsilon_{ijk} ds_k$ by application of Eq. 3.

In differential form, the surface area is naturally expressed as an anti-symmetric second-rank tensor $d\underline{A}$:

$$dA_{ij} = J_{ij} du_1 du_2 \tag{14}$$

where $\{u_1, u_2\}$ are the two independent variables needed to describe a surface $g(x_1, x_2, x_3) \equiv g(u_1, u_2)$ in three-dimensional space, and $\{J_{ij}\}$ are the nine components – of which 3 are zero and only three are independent – corresponding to the Jacobian transformation of $(x_i, x_j)$ with respect to $(u_1, u_2)$:

$$J_{ij} = \begin{vmatrix} \dfrac{\partial x_i}{\partial u_1} & \dfrac{\partial x_i}{\partial u_2} \\ \dfrac{\partial x_j}{\partial u_1} & \dfrac{\partial x_j}{\partial u_2} \end{vmatrix} \tag{15}$$

In rectangular coordinates, the differential area tensor assumes the simple form $dA_{ij} = dx_i dx_j$. The differential area also assumes a vector form with the help of Eq. 3: $dA_k = \dfrac{1}{2!} \varepsilon_{ijk} dA_{ij}$. The scalar form of the differential area is $dA = \sqrt{dA_i dA_i}$.

When generalizing to higher dimensions, we find it convenient to work with the Jacobians because the antisymmetry associated with row and column interchanges in the determinant makes it easy to identify the transformation properties. The same can be done for the differential line element, where just one independent variable – i.e. the arc length $s$ – is needed to describe the distance traveled along a curve $f(x_1, x_2, x_3) \equiv f(s)$ winding through three-dimensional space: Namely, $ds_i = I_i ds$, where $I_i = \partial x_i / \partial s$ is the $1 \times 1$ Jacobian transformation from $x_i$ to $s$. Only the scalar form of the differential volume element is significant in three-dimensional space: $dV = dx_1 dx_2 dx_3$.

In four-dimensional space, the hypervolume is simply $dW = dx_1 dx_2 dx_3 dx_4$ and the other differential elements assume the form:

$$\begin{aligned}
ds &= \sqrt{ds_I ds_I} & dA &= \sqrt{\frac{dA_{IJ} dA_{IJ}}{2!}} & dV &= \sqrt{dV_I dV_I} \\
ds_I &= dx_I & & & dV_L &= (1/3!)\varepsilon_{IJKL} K_{IJK} dv_1 dv_2 dv_3 \quad (16) \\
ds_{IJK} &= \varepsilon_{IJKL} dx_L & dA_{IJ} &= J_{IJ} du_1 du_2 & dV_{IJK} &= K_{IJK} dv_1 dv_2 dv_3
\end{aligned}$$

where $g(x_1, x_2, x_3, x_4) \equiv g(u_1, u_2)$ and $h(x_1, x_2, x_3, x_4) \equiv h(v_1, v_2, v_3)$ describe two- and three-dimensional surfaces in four-dimensional space, respectively, and $\{K_{IJK}\}$ are the 64 components – of which 24 are nonzero and only 4 are independent – corresponding to the Jacobian transformation of $(x_I, x_J, x_K)$ with respect to $(v_1, v_2, v_3)$:

$$K_{IJK} = \begin{vmatrix} \frac{\partial x_I}{\partial v_1} & \frac{\partial x_I}{\partial v_2} & \frac{\partial x_I}{\partial v_3} \\ \frac{\partial x_J}{\partial v_1} & \frac{\partial x_J}{\partial v_2} & \frac{\partial x_J}{\partial v_3} \\ \frac{\partial x_K}{\partial v_1} & \frac{\partial x_K}{\partial v_2} & \frac{\partial x_K}{\partial v_3} \end{vmatrix} \quad (17)$$

In four-dimensional space, the components $\{J_{IJ}\}$ form an anti-symmetric second-rank $4 \times 4$ tensor with 6 independent components. In five-dimensional space,

$$\begin{aligned}
ds &= \sqrt{ds_{I_5} ds_{I_5}} & dV &= \sqrt{dV_{I_5} dV_{I_5}} \\
ds_{I_5} &= dx_{I_5} & dV_{I_5 J_5} &= \frac{1}{3!}\varepsilon_{I_5 J_5 K_5 L_5 M_5} K_{K_5 L_5 M_5} dv_1 dv_2 dv_3 \\
ds_{I_5 J_5 K_5 L_5} &= \varepsilon_{I_5 J_5 K_5 L_5 M_5} dx_{M_5} & dV_{I_5 J_5 K_5} &= K_{I_5 J_5 K_5} dv_1 dv_2 dv_3 \quad (18) \\
dA &= \sqrt{\frac{dA_{I_5 J_5} dA_{I_5 J_5}}{2!}} & dW &= \sqrt{dW_{I_5} dW_{I_5}} \\
dA_{I_5 J_5} &= J_{I_5 J_5} du_1 du_2 & dW_{I_5} &= \frac{1}{4!}\varepsilon_{I_5 J_5 K_5 L_5 M_5} L_{J_5 K_5 L_5 M_5} dw_1 dw_2 dw_3 dw_4 \\
dA_{I_5 J_5 K_5} &= \frac{1}{2!}\varepsilon_{I_5 J_5 K_5 L_5 M_5} J_{L_5 M_5} du_1 du_2 & dW_{I_5 J_5 K_5 L_5} &= L_{I_5 J_5 K_5 L_5} dw_1 dw_2 dw_3 dw_4
\end{aligned}$$

where each quantity in Eq. 16 has been generalized to five-dimensional space.

In $(N+1)$-D spacetime, Gauss's law and Stokes's theorem involve higher-dimensional analogs of the differential line element $d\vec{s}$, differential surface area tensor $d\underline{\mathbf{A}}$, and scalar differential volume $dV$. The higher-dimensional analog of the differential line element $d\vec{s}$ will have $N-2$ spatial dimensions, will be most naturally expressed as an anti-symmetric $(N-2)$-rank tensor in terms of an $(N-2)$-D Jacobian, and can be re-expressed as an anti-symmetric second-rank tensor via an $N$-dimensional Levi-Civita symbol. The higher-dimensional analog of the differential area tensor $d\underline{\mathbf{A}}$ will have $N-1$ spatial dimensions, will be most naturally expressed as an anti-symmetric $(N-1)$-rank tensor in terms of an $(N-1)$-D Jacobian, and can be re-expressed as a vector via an $N$-dimensional Levi-Civita symbol. The higher-dimensional analog of the scalar differential volume $dV$ is simply $dx_1 dx_2 \cdots dx_N$ in rectangular coordinates. (Of course, the differential line element $d\vec{s}$ itself will naturally be a vector $ds_{I_N} = dx_{I_N}$ and the differential area $d\underline{\mathbf{A}}$ will still naturally be an anti-symmetric second-rank tensor for any value of $N$. However, Gauss's law and Stokes's theorem in higher dimensions do not involve the differential line element $d\vec{s}$ and the differential area $d\underline{\mathbf{A}}$ themselves, but, rather, their higher-dimensional analogs.)

**4. Maxwell's Equations**

We work in SI units in order to illustrate how the permittivity and permeability of free space, $\varepsilon_0$ and $\mu_0$, respectively, are altered in higher dimensions. In $(3+1)$-D, the field strength tensor $\underline{\mathbf{F}}$ can be represented as an anti-symmetric $4 \times 4$ matrix:

$$\underline{\mathbf{F}} = \begin{pmatrix} 0 & F^{01} & F^{02} & F^{03} \\ -F^{01} & 0 & F^{12} & F^{13} \\ -F^{02} & -F^{12} & 0 & F^{23} \\ -F^{03} & -F^{13} & -F^{23} & 0 \end{pmatrix} \quad (19)$$

We define the contravariant spacetime position vector as $\{x^\mu\} \equiv \{x^0 \equiv ct, x^1, x^2, x^3\}$, where the lowercase Greek letters represent spacetime indices $\{\mu, \nu\} = \{0,1,2,3\}$. We choose the metric tensor $\underline{\mathbf{g}} \equiv \text{diag}\{-1,1,1,1\}$ to have positive spatial components such that the raising and lowering of indices only changes the sign of the temporal components. The components of the field strength tensor $\{F^{\mu\nu}\}$ are related to the components of the spacetime potential $\{A^\mu\} = \{A^0 \equiv V/c, A^1, A^2, A^3\}$ via

$$F^{\mu\nu} = \partial^\mu A^\nu - \partial^\nu A^\mu \quad (20)$$

The field strength tensor $\underline{\mathbf{F}}$ is naturally divided into temporal and spatial components. The three independent temporal components are associated with the vector electric field

$$E^i/c \equiv F^{0i} = \partial^0 A^i - \partial^i A^0 \qquad (21)$$

while the three spatial components naturally form a $3 \times 3$ second-rank anti-symmetric tensor magnetic field

$$B^{ij} \equiv F^{ij} = \partial^i A^j - \partial^j A^i \qquad (22)$$

although the three independent components of $\underline{\mathbf{B}}$ are usually adapted to form a pseudovector according to Eq. 3. We emphasize that this more natural tensor form generalizes to higher dimensions, while it is merely coincidental that the magnetic field may be interpreted as a pseudovector in $(3+1)$-D.

Maxwell's inhomogeneous equations, in vacuum, are expressed concisely in terms of the field strength tensor:

$$\partial_\nu F^{\mu\nu} = \mu_0 J^\mu \qquad (23)$$

where the spacetime current density has components $J^\mu = \{J^0 \equiv c\rho, J^1, J^2, J^3\}$. In $(3+1)$-D, $\rho$ is the volume charge density and $\vec{\mathbf{J}} = J_1 \hat{\mathbf{x}}_1 + J_2 \hat{\mathbf{x}}_2 + J_3 \hat{\mathbf{x}}_3$ represents the distribution of the current over a cross-sectional area. Maxwell's homogeneous equations follow from the relationship between the fields and the spacetime potential (Eq.'s 21-22). In terms of the vector electric field $\vec{\mathbf{E}}$ and tensor magnetic field $\underline{\mathbf{B}}$, Maxwell's equations in differential form are:

$$\begin{aligned} \partial_i E_i &= \frac{\rho}{\varepsilon_0} & \partial_i E_j - \partial_j E_i &= -c\partial_0 B_{ij} \\ \varepsilon_{ijk} \partial_i B_{jk} &= 0 & -\partial_i B_{ij} &= \frac{\partial_0 E_j}{c} + \mu_0 J_j \end{aligned} \qquad (24)$$

We do not distinguish between covariant and contravariant spatial indices $\{i,j,k\} \in \{1,2,3\}$ since only raising or lowering of temporal components involves a sign change. However, it is crucial to note that $\partial^0$ and $\partial_0$ differ in sign. Note, for example, that the components $E^i/c = \partial^0 A^i - \partial^i A^0$ and $E_i/c = -\partial_0 A_i + \partial_i A_0$ are numerically identical. With the line element $d\underline{\mathbf{s}}$ expressed as an anti-symmetric second-rank tensor and the differential area $d\vec{\mathbf{A}}$ expressed as a vector (via Eq. 3), the integral form of Maxwell's equations are:

$$\oint_S E_i dA_i = \frac{1}{\varepsilon_0} \int_{V_{enc}} \rho dV \qquad \varepsilon_{ijk} \oint_C E_i ds_{jk} = -c\varepsilon_{ijk} \partial_0 \int_{S_{enc}} B_{ij} dA_k$$
$$\varepsilon_{ijk} \oint_S B_{ij} dA_k = 0 \qquad \oint_C B_{ij} ds_{ij} = 2! \int_{S_{enc}} \left( \mu_0 J_i + \frac{\partial_0 E_i}{c} \right) dA_i \qquad (25)$$

When generalizing Eq. 25 to $(N+1)$-D, the higher-dimensional analog of the differential line element $d\underline{\mathbf{s}}$ will have $N-2$ spatial dimensions and the higher-dimensional analog of the differential area $d\vec{\mathbf{A}}$ will have $N-1$ spatial dimensions.

It is also useful to work with an anti-symmetric second-rank tensor $\underline{\mathbf{G}}$ that is dual to $\underline{\mathbf{F}}$:[§]

$$G^{\mu\nu} \equiv \frac{1}{2!} \varepsilon^{\mu\nu\rho\sigma} F_{\rho\sigma} \qquad (26)$$

Maxwell's homogeneous equations are obtained from the dual tensor $\underline{\mathbf{G}}$ via

$$\partial_\nu G^{\mu\nu} = 0 \qquad (27)$$

Alternatively, Maxwell's homogeneous equations may be extracted from

$$\varepsilon^{\lambda\mu\nu\rho} T_{\lambda\mu\nu} = 0 \qquad (28)$$

where the anti-symmetric third-rank tensor $\underline{\mathbf{T}}$ is defined as

$$T_{\lambda\mu\nu} \equiv \partial_\lambda F_{\mu\nu} \qquad (29)$$

The most straightforward generalization of Maxwell's equations to $(4+1)$-D begins by extending Eq. 21 to

$$\partial_\Gamma F^{\Phi\Gamma} = \mu_{4+1} J^\Phi \qquad (30)$$

where the uppercase Greek letters $\{\Phi, \Gamma\} \in \{0,1,2,3,4\}$ represent $(4+1)$-D spacetime indices and the $(4+1)$-D permittivity and permeability of free space $\varepsilon_{4+1}$ and $\mu_{4+1}$, respectively, have different dimensions than their $(3+1)$-D counterparts $\varepsilon_0$ and $\mu_0$. The $5 \times 5$ $(4+1)$-D field strength tensor $\underline{\mathbf{F}}$ has 10 independent components, which naturally divides into a 4-component vector electric field

$$E^I/c = F^{0I} = \partial^0 A^I - \partial^I A^0 \qquad (31)$$

and a $4 \times 4$ anti-symmetric second rank tensor magnetic field

$$B^{IJ} = F^{IJ} = \partial^I A^J - \partial^J A^I \qquad (32)$$

with 6 independent components. In terms of the vector electric field $\vec{\mathbf{E}}$ and tensor magnetic field $\underline{\mathbf{B}}$, Maxwell's equations in $(4+1)$-D in differential form become

$$\begin{aligned}\partial_I E_I &= \frac{\rho}{\varepsilon_{4+1}} & \partial_I E_J - \partial_J E_I &= -c\partial_0 B_{IJ} \\ \varepsilon_{IJKL}\partial_I B_{JK} &= 0 & -\partial_I B_{IJ} &= \frac{\partial_0 E_J}{c} + \mu_{4+1} J_J\end{aligned} \qquad (33)$$

---

[§] We adhere to the convention that the contravariant form of the $(3+1)$-D spacetime Levi-Civita symbol $\varepsilon^{\mu\nu\rho\sigma}$ be based upon $\varepsilon^{0123} = -1$ such that $\varepsilon_{0123} = 1$, not to be confused with the Levi-Civita symbol $\varepsilon_{IJKL}$ corresponding to four spatial dimensions, in which $\varepsilon_{1234} = 1$. Again, in $(4+1)$-D, we choose $\varepsilon^{\Phi\Gamma\Lambda\Sigma\Omega}$ to be based upon $\varepsilon^{01234} = -1$.

Gauss's law in magnetism, which states that the net magnetic flux is always zero as a consequence that magnetic monopoles have not been observed, can alternatively be expressed in differential form as:

$$\partial_I B_{JK} + \partial_J B_{KI} + \partial_K B_{IJ} = 0 \qquad (34)$$

Other formulas, which follow, involving the higher-dimensional Levi-Civita symbol may be expressed in this form (Eq. 34) via Eq. 11 or its higher-dimensional analogs.

Generalization of the integral form of Maxwell's equations is more subtle. In four spatial dimensions, the differential elements in Gauss's law and Stokes's theorem also grow in dimension: line element becomes differential area, differential area becomes differential volume, and differential volume becomes differential hypervolume. The differential area $d\underline{\mathbf{A}}$ naturally forms an anti-symmetric second-rank tensor and the differential volume $d\underline{\mathbf{V}}$ naturally forms an anti-symmetric third-rank tensor. Alternatively, the differential volume may also be expressed as a vector $d\vec{\mathbf{V}}$ via Eq. 9. With the fields and differential elements expressed in terms of appropriate vectors and tensors, $(4+1)$-D Maxwell's equations can be expressed in integral form as:[**]

$$\begin{aligned}\oint_V E_I dV_I &= \frac{1}{\varepsilon_{4+1}} \int_{W_{enc}} \rho dW & \varepsilon_{IJKL}\oint_S E_I dA_{JK} &= -c\varepsilon_{IJKL}\partial_0 \int_{V_{enc}} B_{IJ} dV_K \\ \varepsilon_{IJKL}\oint_V B_{IJ} dV_K &= 0 & \oint_S B_{IJ} dA_{IJ} &= 2! \int_{V_{enc}}\left(\mu_{4+1} J_I + \frac{\partial_0 E_I}{c}\right)dV_I\end{aligned} \qquad (35)$$

---

[**] In this section, we assume the extra dimensions to be of the same noncompact nature as the usual three spatial dimensions. We consider the effects of compactification in Sec.'s 6-7.

While electric flux remains a scalar, magnetic flux has four components in $(4+1)$-D:

$$\Phi^e = \int_V E_I dV_I$$
$$\Phi^m_L = \varepsilon_{IJKL} \int_V B_{IJ} dV_K \tag{36}$$

In $(4+1)$-D, the continuity equation reads

$$\partial_I J_I = -c\partial_0 \rho \tag{37}$$

In $(4+1)$-D, the dual tensor $\underline{\underline{G}}$ is an anti-symmetric third-rank tensor:

$$G^{\Phi\Gamma\Lambda} \equiv \frac{1}{3!} \varepsilon^{\Phi\Gamma\Lambda\Sigma\Omega} F_{\Sigma\Omega} \tag{38}$$

Maxwell's homogeneous equations may be expressed in terms of the dual tensor $\underline{\underline{G}}$ via

$$\partial_\Lambda G^{\Phi\Gamma\Lambda} = 0 \tag{39}$$

or, equivalently, in terms of an anti-symmetric third-rank tensor $\underline{\underline{T}}$ via

$$\varepsilon^{\Phi\Gamma\Lambda\Sigma\Omega} T_{\Lambda\Sigma\Omega} = 0 \tag{40}$$

where $\underline{\underline{T}}$ is defined as

$$T_{\Phi\Gamma\Lambda} \equiv \partial_\Phi F_{\Gamma\Lambda} \tag{41}$$

In $(5+1)$-D, the differential form of Maxwell's inhomogeneous equations are expressed concisely in terms of the $6\times 6$ field strength tensor $\underline{F}$ as

$$\partial_{\Gamma_{5+1}} F^{\Phi_{5+1}\Gamma_{5+1}} = \mu_{5+1} J^{\Phi_{5+1}} \tag{42}$$

where $\{\Phi_{5+1}, \Gamma_{5+1}\} \in \{0,1,2,3,4,5\}$ represent $(5+1)$-D spacetime indices. Maxwell's homogeneous equations can be expressed in terms of the dual tensor $\underline{\underline{G}}$ or, equivalently, its counterpart $\underline{\underline{T}}$, as

$$\partial_{\Xi_{5+1}} G^{\Phi_{5+1}\Gamma_{5+1}\Lambda_{5+1}\Xi_{5+1}} = 0$$
$$\varepsilon^{\Phi_{5+1}\Gamma_{5+1}\Lambda_{5+1}\Sigma_{5+1}\Omega_{5+1}\Xi_{5+1}} T_{\Lambda_{5+1}\Sigma_{5+1}\Omega_{5+1}} = 0 \tag{43}$$

where

$$G^{\Phi_{5+1}\Gamma_{5+1}\Lambda_{5+1}\Sigma_{5+1}} \equiv \frac{1}{4!}\varepsilon^{\Phi_{5+1}\Gamma_{5+1}\Lambda_{5+1}\Sigma_{5+1}\Omega_{5+1}\Xi_{5+1}} F_{\Omega_{5+1}\Xi_{5+1}}$$ (44)

$$T_{\Phi_{5+1}\Gamma_{5+1}\Lambda_{5+1}} \equiv \partial_{\Phi_{5+1}} F_{\Gamma_{5+1}\Lambda_{5+1}}$$

In terms of the 5-component vector electric field $\vec{\mathbf{E}}$ and the $5\times 5$ tensor magnetic field $\underline{\mathbf{B}}$, the differential form of Maxwell's equations in $(5+1)$-D are:

$$\partial_{I_5} E_{I_5} = \frac{\rho}{\varepsilon_{5+1}} \qquad \partial_{I_5} E_{J_5} - \partial_{J_5} E_{I_5} = -c\partial_0 B_{I_5 J_5}$$

$$\varepsilon_{I_5 J_5 K_5 L_5 M_5} \partial_{I_5} B_{J_5 K_5} = 0 \qquad -\partial_{I_5} B_{I_5 J_5} = \frac{\partial_0 E_{J_5}}{c} + \mu_{5+1} J_{J_5}$$ (45)

where $\{I_5, J_5\} = \{1,2,3,4,5\}$. In integral form, Maxwell's equations in $(5+1)$-D are:

$$\oint_W E_{I_5} dW_{I_5} = \frac{1}{\varepsilon_{5+1}} \int_{X_{enc}} \rho\, dX \qquad \varepsilon_{I_5 J_5 K_5 L_5 M_5} \oint_V E_{I_5} dV_{J_5 K_5} = -c\varepsilon_{I_5 J_5 K_5 L_5 M_5} \partial_0 \int_{W_{enc}} B_{I_5 J_5} dW_{K_5}$$

$$\varepsilon_{I_5 J_5 K_5 L_5 M_5} \oint_W B_{I_5 J_5} dW_{K_5} = 0 \qquad \oint_V B_{I_5 J_5} dV_{I_5 J_5} = 2! \int_{W_{enc}} \left(\mu_{5+1} J_{I_5} + \frac{\partial_0 E_{I_5}}{c}\right) dW_{I_5}$$ (46)

where $dX$ is the differential hyperhypervolume (i.e. five-dimensional volume) element.

It is straightforward to extend Maxwell's equations to yet higher dimensions. The $(N+1)$-D analogs of Eq.'s 41-43 involve replacing the $(5+1)$-D spacetime indices $\{\Phi_{5+1}, \Gamma_{5+1}\} \in \{0,1,2,3,4,5\}$ with $(N+1)$-D spacetime indices $\{\Phi_{N+1}, \Gamma_{N+1}\} \in \{0,1,\ldots,N\}$, accommodating the higher dimensionality of $\varepsilon_{N+1}$ and $\mu_{N+1}$, and extending the Levi-Civita spacetime symbol to have $(N+1)$ indices $\varepsilon^{\Phi_{N+1}\Gamma_{N+1}\cdots\Psi_{N+1}}$. In terms of the electric and magnetic fields, the $(N+1)$-D analog of Eq. 45 involves similar replacements, except that these equations instead involve the spatial Levi-Civita symbol with $N$ indices $\varepsilon_{I_N J_N \cdots Z_N}$. The $(N+1)$-D version of the integral form of Maxwell's equations, which is the $(N+1)$-D analog of Eq. 46, also involves the higher-dimensional analogs of the usual line elements $d\vec{\mathbf{s}}$, $d\underline{\mathbf{A}}$, and $dV$, which are, respectively, a differential anti-symmetric second-rank tensor $d^{N-2}\underline{\mathbf{x}}$ with $N-2$ spatial dimensions, a differential vector $d^{N-1}\vec{\mathbf{x}}$ with $N-1$ spatial dimensions, and a differential scalar $d^N x = x_1 x_2 \cdots x_N$ with $N$ spatial dimensions. In addition, some permutation factors may depend on the dimensionality of the spacetime (e.g. compare Eq.'s 11 and 13).

## 5. Force Laws

For an isolated point charge in $(4+1)$-D, the electric field lines radiate outward in four-dimensional space. For a Gaussian hypersphere of radius $r$, the electric field $\vec{E}$ is parallel to the differential volume element $d\vec{V}$ everywhere on the three-dimensional volume $x_1^2 + x_2^2 + x_3^2 + x_4^2 = r^2$ bounding the (four-dimensional) hypersphere defined by $x_1^2 + x_2^2 + x_3^2 + x_4^2 \leq r^2$. In $(4+1)$-D, the electric field $\vec{E}$ due to a point charge is thus found to be[††]

$$\oint_V E_I dV_I = E 2\pi^2 r^3 = \frac{1}{\varepsilon_{4+1}} \int_{W_{enc}} \rho dW = \frac{q}{\varepsilon_{4+1}}$$

$$\vec{E} = \frac{q}{2\pi^2 \varepsilon_{4+1} r^3} \hat{r} \tag{47}$$

We have utilized the result that the $(N-1)$-dimensional volume $\sum_{i=1}^{N} x_i^2 = r^2$ bounding a solid sphere in $N$-dimensional space $\sum_{i=1}^{N} x_i^2 \leq r^2$ is

$$V_{N-1}^{sphere} = \frac{2\pi^{N/2} r^{N-1}}{\Gamma(N/2)} \tag{48}$$

where $\Gamma(z)$ is the gamma function. For example, the one-dimensional volume (i.e. circle) bounding the two-dimensional solid sphere (i.e. disk) is $2\pi r$, while the two-dimensional volume (i.e. spherical surface) bounding the three-dimensional solid sphere is $4\pi r^2$.

Generalizing to $(N+1)$-D, the electric field $\vec{E}$ due to a point charge is

$$\vec{E} = \frac{\Gamma(N/2)}{2\pi^{N/2}} \frac{q}{\varepsilon_{N+1} r^{N-1}} \hat{r} \tag{49}$$

where the $(N+1)$-D permittivity of free space $\varepsilon_{N+1}$ has different dimensionality than its $(3+1)$-D counterpart $\varepsilon_0$:[‡‡]

$$[\varepsilon_{N+1}] = \frac{[\varepsilon_0]}{[L^{N-3}]} \tag{50}$$

---

[††] In this section, we assume the extra dimensions to be of the same noncompact nature as the usual three spatial dimensions. We consider the effects of compactification in Sec.'s 6-7.

[‡‡] We relate the $(3+1)$-D and $(N+1)$-D permittivites in Sec. 7 for the case of toroidal compactification.

Thus, Coulomb's law for the force exerted by one point charge $q_1$ on another point charge $q_2$ displaced by the relative position vector $\vec{R}_{21}$ away from the first is a $1/r^{N-1}$ force law:

$$\vec{F}_{q_2,q_1} = \frac{\Gamma(N/2)}{2\pi^{N/2}} \frac{q_1 q_2}{\varepsilon_{N+1} R_{21}^{N-1}} \hat{R}_{21} \tag{51}$$

In $(N+1)$-D, the electric field $\vec{E}(\vec{r})$ at a field point located at $\vec{r}$ may alternatively be derived through direct integration over the differential source element $dq'$:

$$\vec{E}(\vec{r}) = \frac{\Gamma(N/2)}{2\pi^{N/2} \varepsilon_{N+1}} \int_0^{q'} \frac{\hat{R} dq'}{R^{N-1}} \tag{52}$$

where the relative position vector $\vec{R} = \vec{r} - \vec{r}'$ extends from the differential source element $dq'$ located at $\vec{r}'$ to the field point located at $\vec{r}$. Depending upon the dimensionality of the source, the differential source element $dq'$ may be expressed as: $\lambda ds$ for a one-dimensional arc, $\sigma dA$ for a two-dimensional surface, $\rho_3 dV$ for a three-dimensional volume, $\rho_4 dW$ for a four-dimensional hypervolume, etc. The $(N+1)$-D electric field is related to the $(N+1)$-D scalar potential via

$$\vec{E} = -\vec{\nabla} V - \frac{\partial \vec{A}}{\partial t} \tag{53}$$

In electrostatics, the $(N+1)$-D scalar potential is derived from the following integral:

$$V(r) = \frac{\Gamma(N/2)}{2\pi^{N/2}(N-2)\varepsilon_{N+1}} \int_0^{q'} \frac{dq'}{R^{N-2}} \tag{54}$$

Consider an infinite straight filamentary conductor running along the $x_1$-axis. In $(3+1)$-D, the magnetic flux $\Phi^m$ is a scalar and the magnetic field lines are traditionally drawn as concentric circles in the $x_2 x_3$ plane. A convenient Ampèrian loop is therefore a circle lying in the $x_2 x_3$ plane. However, in $(4+1)$-D, magnetic flux is a four-component vector (Eq. 36) and any vector in the three-dimensional space $x_2 x_3 x_4$ is orthogonal to the current running along the $x_1$-axis. The traditional Ampèrian loop is promoted to a two-dimensional Ampèrian surface in $(4+1)$-D. A suitable choice for a steady filamentary current $I_0$ running along the $x_1$-axis is the two-dimensional surface $x_2^2 + x_3^2 + x_4^2 = r^2$ of a solid sphere $x_2^2 + x_3^2 + x_4^2 \leq r^2$ in the three-dimensional subspace $x_2 x_3 x_4$. In this case, the only non-vanishing components of the magnetic field tensor $\underline{\mathbf{B}}$ are $\{B_{1I}\}$ and $\{B_{I1}\}$. By symmetry, the magnetic field scalar, defined by the tensor contraction $B \equiv \sqrt{B_{IJ}B_{IJ}/2!}$, is constant everywhere on the surface $x_2^2 + x_3^2 + x_4^2 = r^2$ and the components of the magnetic field tensor are related by $B_{II} = B_{IJ} = -B_{I1} = -B_{J1}$, where $I \neq J$, such that $B = \sqrt{B_{12}^2 + B_{13}^2 + B_{14}^2} = B_{12}\sqrt{3}$. However, it is simpler to work in spherical coordinates, where $B_{IJ}dA_{IJ} = 2Br^2 d\Omega$. Thus, application of Ampère's law yields

$$\oint_S B_{IJ} dA_{IJ} = 2\int_S Br^2 d\Omega = B8\pi r^2 = 2\mu_{4+1}I_{enc}$$

$$B = \frac{\mu_{4+1}I_0}{4\pi r^2} \tag{55}$$

In $(3+1)$-D, the magnetic flux $\Phi^m$ is a scalar and the magnetic field lines for this infinite steady filamentary current $I_0$ are traditionally drawn as concentric circles $x_2^2 + x_3^2 = r^2$ in the $x_2 x_3$ plane. However, in $(4+1)$-D the magnetic flux is a four-component vector $\{\Phi_I^m\}$ according to Eq. 36. This corresponds to the fact that circles $x_2^2 + x_3^2 = r^2$, $x_2^2 + x_4^2 = r^2$, and $x_3^2 + x_4^2 = r^2$ lying in the $x_2 x_3$, $x_2 x_4$, and $x_3 x_4$ planes, respectively, are all orthogonal to the current running along the $x_1$-axis. In $(N+1)$-D, the magnetic flux becomes an anti-symmetric $(N-3)$-rank tensor.

Generalizing to $(N+1)$-D, the magnetic field due to a steady filamentary current $I_0$ is

$$B(r) = \frac{\Gamma\left(\frac{N-1}{2}\right)}{2\pi^{(N-1)/2}} \frac{\mu_{4+1}I_0}{r^{N-2}} \tag{56}$$

In $(N+1)$-D, the force per unit length $\ell$ that one steady filamentary current $I_1$ exerts on a parallel steady filamentary current $I_2$ separated by the relative position vector $\vec{R}_{21}$ is:

$$\frac{\vec{F}_{I_2,I_1}}{\ell} = -\frac{\Gamma\left(\frac{N-1}{2}\right)}{2\pi^{(N-1)/2}} \frac{\mu_{4+1}I_1 I_2}{R_{21}^{N-2}} \hat{R}_{21} \tag{57}$$

In $(4+1)$-D, the tensor magnetic field $\underline{B}(\vec{r})$ at a field point located at $\vec{r}$ may alternatively be derived through direct integration. For a steady filamentary current,

$$B_{IJ}(\vec{r}) = \frac{\Gamma\left(\frac{N-1}{2}\right)}{4\pi^{(N-1)/2}} \mu_{N+1}I_0 \int \frac{R_J ds_I - R_I ds_J}{R^4} \tag{58}$$

where the relative position vector $\vec{R} = \vec{r} - \vec{r}'$ extends from the differential source element $d\vec{s}'$ located at $\vec{r}'$ to the field point located at $\vec{r}$. If the current is instead distributed over a one-, two-, or three-dimensional cross section, the tensor magnetic field $\underline{\mathbf{B}}(\vec{r})$ is

$$B_{IJ}(\vec{r}) = \frac{\Gamma\left(\frac{N-1}{2}\right)}{4\pi^{(N-1)/2}} \mu_{N+1} \int \frac{J_1 dA_{IJ}}{R^3} \qquad B_{IJ}(\vec{r}) = \frac{\Gamma\left(\frac{N-1}{2}\right)}{4\pi^{(N-1)/2}} \mu_{N+1} \int \frac{J_2 dA_{IJ} ds}{R^3}$$

$$B_{IJ}(\vec{r}) = \frac{\Gamma\left(\frac{N-1}{2}\right)}{4\pi^{(N-1)/2}} \mu_{N+1} \int \frac{J_3 (R_J dV_I - R_I dV_J) ds}{R^4}$$

(59)

where $J_1$, $J_2$, and $J_3$ are the corresponding the current densities. Here, the differential arc length $d\vec{s}$ is a vector along $I$. Note that the differential area element $d\underline{\mathbf{A}}$ associated with $J_1$ is comprised of $d\vec{s}$ and the differential arc length $d\vec{w}$ over which the current is distributed, while the differential area element $d\underline{\mathbf{A}}$ associated with $J_2$ does not – rather, it corresponds to the area over which the current is distributed. Similarly, the differential volume element $d\vec{V}$ associated with $J_3$ corresponds to the volume over which the current is distributed, and therefore does not include $d\vec{s}$.

The case for magnetic monopoles is not as compelling in higher dimensions as it is in $(3+1)$-D. Maxwell's equations are conventionally expressed very symmetrically in $(3+1)$-D, in which case the electric and magnetic fields each have three independent components. In $(3+1)$-D, Ampère's law and Gauss's law in magnetism have almost the exact same form as Faraday's law and Gauss's law in electricity, respectively. It is very tempting, aesthetically, to introduce magnetic monopoles into the theory in order to complete this symmetry. However, in higher dimensions, the magnetic field behaves much differently than the electric field: The magnetic field is an anti-symemtric second-rank tensor $\underline{\mathbf{B}}$, with more components than the vector electric field $\vec{\mathbf{E}}$. Thus, in higher dimensions, the equations are not as symmetric between the electric and magnetic fields. In $(3+1)$-D, the apparent symmetry observed in the dual tensor $\underline{\mathbf{G}}$ – i.e. $\vec{\mathbf{E}}/c \to \vec{\mathbf{B}}$ and $\vec{\mathbf{B}} \to \vec{\mathbf{E}}/c$ – is accidental.

## 6. Compactification

The most straightforward generalization of the usual $(3+1)$-D SM Lagrangian density $L_{SM}(x^\mu)$ for the gauge fields to $(4+1)$-D is[§§]

---

[§§] In this section only, we work in Heaviside-Lorentz units, setting $\hbar = c = 1$, as this context is more appropriate for this high-energy physics content.

$$L_{4+1}(x^\Lambda) = -\frac{F_{\Phi\Gamma} F^{\Phi\Gamma}}{4} - \sum_f J_{f\Phi}(x^\Lambda) A^\Phi(x^\Lambda) \tag{60}$$

$J_{f\Phi}(x^\Lambda) = q_f \overline{\psi}_f(x^\Lambda) \Gamma_\Phi \psi_f(x^\Lambda)$ is the $(4+1)$-D electromagnetic current corresponding to the fermionic multiplet $\psi_f(x^\Lambda)$ and $\{\Gamma^\Phi\}$ are the $(4+1)$-D generalizations of the $(3+1)$-D Dirac gamma matrices $\{\gamma^\mu\}$. Everyday experience indicates that the extra dimension $x^4$ is not macroscopically observable. This problem is easily solved if the extra dimension $x^4$ is compactified on an $S^1/Z_2$ orbifold with radius $R$ that is microscopic – in particular, the compactification scale $1/R$ must exceed ~350-400 GeV for the KK excitations of the fields not to have already been observed at high-energy colliders such as the Tevatron. The $Z_2$ parity insures that each component of the $(4+1)$-D fields is either even or odd under the orbifold transformation $x^4 \to -x^4$ such that the effective $(3+1)$-D theory will contain zero modes of the fields which are normally observed in the SM (even $Z_2$ parity), while the zero modes corresponding to fields that are not observed in the SM can be projected out (odd $Z_2$ parity). KK number and $Z_2$ parity are conserved in the interactions.

The components of the $(4+1)$-D gauge field $A^\Phi(x^\Lambda)$ can be expanded in a Fourier series about the compactified dimension $x^4$ as

$$A^\mu(x^\nu, x^4) = \frac{1}{\sqrt{\pi R}} \left[ A^{\mu(0)}(x^\nu) + \sqrt{2} \sum_{n=1}^{\infty} A^{\mu(n)}(x^\nu) \cos\left(\frac{nx^4}{R}\right) \right]$$
$$A^4(x^\nu, x^4) = \sqrt{\frac{2}{\pi R}} \sum_{n=1}^{\infty} A^{\mu(n)}(x^\nu) \sin\left(\frac{nx^4}{R}\right) \tag{61}$$

The usual $(3+1)$-D spacetime components of the gauge field $A^\mu(x^\nu, x^4)$ have even $Z_2$ parity $A^\mu(x^\nu, -x^4) = A^\mu(x^\nu, -x^4)$ such that the zero mode $A^{\mu(0)}(x^\nu)$ corresponds to the SM photon, while the component of the gauge field $A^4(x^\nu, x^4)$ polarized along the compact dimension $x^4$ is odd under $Z_2$ parity $A^\mu(x^\nu, -x^4) = -A^\mu(x^\nu, -x^4)$ in order to project its zero mode, which is not observed in the SM, out of the effective $(3+1)$-D theory.

The $(4+1)$-D quark multiplets $Q_g(x^\mu, x^4)$, $U_g(x^\mu, x^4)$, and $D_g(x^\mu, x^4)$ consist of four-component vector-like quark fields, where the generational index $g \in \{1,2,3\}$. The $(4+1)$-D quark multiplets can be decomposed into $(3+1)$-D two-component Weyl spinors as:

$$Q_g(x^\mu,x^4) = \frac{1}{\sqrt{\pi R}}\left\{q_{gL}^{(0)}(x^\mu) + \sqrt{2}\sum_{n=1}^{\infty}\left[q_{gL}^{(n)}(x^\mu)\cos\left(\frac{nx^4}{R}\right) + q_{gR}^{(n)}(x^\mu)\sin\left(\frac{nx^4}{R}\right)\right]\right\}$$

$$U_g(x^\mu,x^4) = \frac{1}{\sqrt{\pi R}}\left\{u_{gR}^{(0)}(x^\mu) + \sqrt{2}\sum_{n=1}^{\infty}\left[u_{gR}^{(n)}(x^\mu)\cos\left(\frac{nx^4}{R}\right) + u_{gL}^{(n)}(x^\mu)\sin\left(\frac{nx^4}{R}\right)\right]\right\} \quad (62)$$

$$D_g(x^\mu,x^4) = \frac{1}{\sqrt{\pi R}}\left\{d_{gR}^{(0)}(x^\mu) + \sqrt{2}\sum_{n=1}^{\infty}\left[d_{gR}^{(n)}(x^\mu)\cos\left(\frac{nx^4}{R}\right) + d_{gL}^{(n)}(x^\mu)\sin\left(\frac{nx^4}{R}\right)\right]\right\}$$

In addition to the usual SM left-chiral doublet $\begin{pmatrix} u_g^{(0)}(x^\mu) \\ d_g^{(0)}(x^\mu) \end{pmatrix}_L$ and right-chiral singlets $u_{gR}^{(0)}(x^\mu)$ and $d_{gR}^{(0)}(x^\mu)$ and their associated KK excitations, there are KK excitations corresponding to a right-chiral doublet $\begin{pmatrix} u_g^{(n)}(x^\mu) \\ d_g^{(n)}(x^\mu) \end{pmatrix}_R$ and left-chiral singlets $u_{gL}^{(n)}(x^\mu)$ and $d_{gL}^{(n)}(x^\mu)$, which are not observed in the SM. There are similar Fourier expansions for the $(4+1)$-D lepton multiplets $L_g(x^\mu,x^4)$, $E_g(x^\mu,x^4)$, and $N_g(x^\mu,x^4)$ corresponding to the SM left-chiral doublet $\begin{pmatrix} e_g^{(0)}(x^\mu) \\ \nu_{eg}^{(0)}(x^\mu) \end{pmatrix}_L$ and right-chiral singlets $e_{gR}^{(0)}(x^\mu)$ and $\nu_{egR}^{(0)}(x^\mu)$.

The effective $(3+1)$-D Lagrangian density $L_{eff}(x^\mu)$ for the gauge fields is obtained by integrating over the compactified coordinate $x^4$. The kinetic terms in the gauge fields are:

$$L_{eff}^{kin}(x^\rho) = -\frac{1}{4}F_{\mu\nu}^{(0)}F^{\mu\nu(0)} - \frac{1}{4}\sum_{n=0}^{\infty}F_{\mu\nu}^{(n)}F^{\mu\nu(n)} - \frac{1}{2}\frac{n^2}{R^2}\sum_{n=0}^{\infty}A_\mu^{(n)}(x^\rho)A^{\mu(n)}(x^\rho)$$
$$-\frac{1}{2}\frac{n^2}{R^2}\sum_{n=1}^{\infty}\partial_\mu A_4^{(n)}(x^\rho)\partial^\mu A^{4(n)}(x^\rho) - \frac{1}{2}\frac{n}{R}\sum_{n=1}^{\infty}\left[A_\mu^{(n)}(x^\rho)\partial^\mu A^{4(n)}(x^\rho) + A^{\mu(n)}(x^\rho)\partial_\mu A_4^{(n)}(x^\rho)\right] \quad (63)$$

The first term is the usual $(3+1)$-D kinetic term, which, when combined with the interaction of the gauge field with the current density, yields the usual $(3+1)$-D Maxwell's equations through the Euler-Lagrange equation. The second term is a corresponding term for the KK excitations. The third term is a mass term for the KK excitations of the usual SM gauge field $A^\mu(x^\nu)$, which have masses $\{m_n = n/R\}$. The fourth term is a corresponding mass term for the KK excitations associated with the component of the gauge field $A^4(x^\nu,x^4)$ polarized along the compact dimension $x^4$, with the same masses. The last sum represents mixing between the KK excitations polarized along $x^\mu$ and $x^4$. The effective $(3+1)$-D Lagrangian density for the interactions between the gauge field and the current density yield the Feynman rules for the couplings of the fermions to the gauge fields [5].

## 7. Coulomb's Law

With a single noncompact extra dimension $x_4$, two $(4+1)$-D electric charges $q_1$ and $q_2$ communicate via the exchange of a $(4+1)$-D photon. Classically,*** there is a single straight-line path with which the $(4+1)$-D photon can reach $q_2$ from another $q_1$. One result of this single noncompact extra dimension $x_4$ is that Coulomb's law becomes a $1/r^3$ force law. If instead the extra dimension $x_4$ is compactified on a torus with radius $R$, there exists an infinite discrete set of paths with which a classical $(4+1)$-D photon could travel from $q_1$ to $q_2$ (Fig. 1). In addition to traveling directly along the shortest path from $q_1$ to $q_2$, a classical $(4+1)$-D photon can also wind one or more times around the extra dimension before reaching the other charge. In the effective $(3+1)$-D theory, the power of $r$ in Coulomb's law is affected by the discrete set of routes available to the classical $(4+1)$-D photon. The separation between the two charges is $R_4 = \sqrt{\Delta x_1^2 + \Delta x_2^2 + \Delta x_3^2 + \Delta x_4^2}$. It is convenient to work with the three-dimensional separation $R_3 = \sqrt{\Delta x_1^2 + \Delta x_2^2 + \Delta x_3^2}$ and the extra-dimensional separation $|\Delta x_4| \le \pi R$ such that $R_4 = \sqrt{R_3^2 + \Delta x_4^2}$. Note that there is really an extra-dimensional circle of radius $R$ at every point in ordinary three-dimensional space – i.e. the cylinder illustrated in Fig. 1 is constructed by orienting the axis of the cylinder along the three-dimensional separation $R_3$.

The net force exerted on $q_2$ is the superposition of the $(4+1)$-D $1/r^3$ Coulomb forces (Eq. 51) associated with each path:

$$\vec{F} = \frac{q_1 q_2}{2\pi^2 \varepsilon_{4+1}} \sum_{n=-\infty}^{\infty} \frac{\hat{\mathbf{R}}_n}{R_n^3} \tag{64}$$

where $R_n = \sqrt{R_3^2 + (\Delta x_4 + 4\pi n R)^2}$ is the path length corresponding to a classical $(4+1)$-D photon winding around the torus $n$ times and $\hat{\mathbf{R}}_n = \frac{R_3 \hat{\mathbf{R}}_3 + \Delta x_4 \hat{\mathbf{x}}_4}{R_n}$ is a unit vector directed from $q_1$ to $q_2$ (and $\hat{\mathbf{R}}_3$ is a unit vector along the axis of the torus). In terms of its three-dimensional and extra-dimensional components, the net force is

$$\vec{F} = \frac{q_1 q_2}{2\pi^2 \varepsilon_{4+1}} \sum_{n=-\infty}^{\infty} \frac{R_3 \hat{\mathbf{R}}_3 + \Delta x_4 \hat{\mathbf{x}}_4}{\left[R_3^2 + (\Delta x_4 + 4\pi n R)^2\right]^2} \tag{65}$$

---

*** As a simple example illustrative of how the compactification impacts the form of Coulomb's law in the effective $(3+1)$-D theory, we consider classical electric charges and photons – i.e. in this first approximation we do not treat the quantum effects.

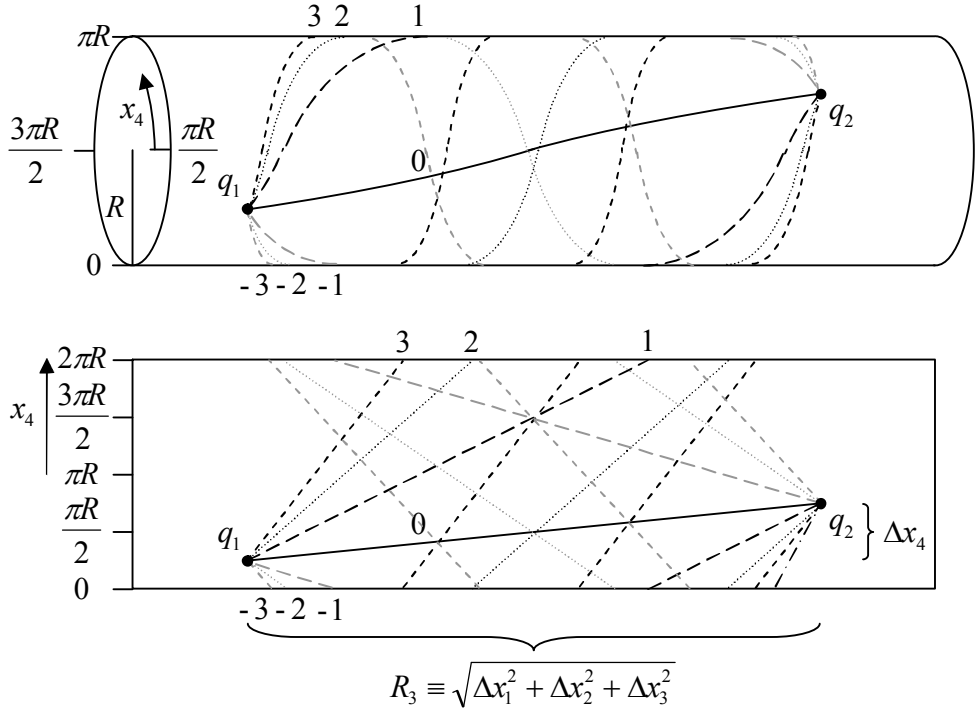

**Fig. 1.** Two stationary classical $(4+1)$-D charges $q_1$ and $q_2$ interact via Coulomb's law with one extra dimension $x_4$ compactified on a torus with radius $R$. A classical $(4+1)$-D photon can reach $q_2$ from $q_1$ (all of which lie on the surface of the torus) by traveling along any of an infinite discrete set of routes, the shortest seven of which are illustrated above. The numerical label indicates the number of times the path winds around the compact dimension, and the minus sign indicates a path (shown in gray) with a negative $x_4$. The bottom figure is identical to the top figure except that the surface of the torus has been unfolded, with the incision made at $x_4 = 0$.

The net force can also be expressed in terms of the $(4+1)$-D scalar potential $V(R_4)$ via

$$\vec{F} = -q_2 \vec{\nabla} V \tag{66}$$

The $(4+1)$-D scalar potential due to the source $q_1$ at the location of $q_2$ is

$$V(R_4) = \frac{q_1}{4\pi^2 \varepsilon_{4+1}} \sum_{n=-\infty}^{\infty} \frac{1}{R_3^2 + (\Delta x_4 + 4\pi nR)^2} \tag{67}$$

Note that the $(4+1)$-D scalar potential is periodic in the extra dimension:

$$V(R_3, \Delta x_4) = V(R_3, \Delta x_4 \pm 2\pi nR) \tag{68}$$

In the limit that the charges are very close compared to the radius of the extra dimension, i.e. $R_3 \ll R$ and $\Delta x_4 \ll R$, the $n=0$ term dominates and the net force is approximately the $(4+1)$-D $1/r^3$ Coulomb force:

$$\vec{F}_{q_1,q_2} \approx \frac{q_1 q_2 \hat{\mathbf{R}}_4}{2\pi^2 \varepsilon_{4+1} R_4^3} \tag{69}$$

This is the limit where the underlying $(4+1)$-D theory of electrodynamics – i.e. Maxwell's equations with a noncompact extra dimension (Sec. 4) – governs the motion. In the opposite extreme, where the compact extra dimension is very small compared to the separation of the charges, i.e. $R_3 \gg R$, an integral provides a good approximation for the sum [11]:

$$V(R_4) \approx \frac{q_1}{8\pi^3 \varepsilon_{4+1} R} \int_{y=-\infty}^{\infty} \frac{dy}{R_3^2 + y^2} = \frac{q_1}{8\pi^2 \varepsilon_{4+1} R R_3} \tag{70}$$

This is the limit where the effective $(3+1)$-D theory of electrodynamics – i.e. the usual $(3+1)$-D form of Maxwell's equations – governs the motion. In order for the effective $(3+1)$-D scalar potential in Eq. 70 to be consistent with the usual $(3+1)$-D form of Coulomb's law, it is necessary that the $(4+1)$-D permittivity of free space $\varepsilon_{4+1}$ be related to the usual $(3+1)$-D permittivity $\varepsilon_0$ via

$$\varepsilon_{4+1} = \frac{\varepsilon_0}{2\pi R} \tag{71}$$

It follows that

$$\mu_{4+1} = 2\pi \mu_0 R \tag{72}$$

such that an electromagnetic wave propagates at the usual speed of light in vacuum

$$c = \frac{1}{\sqrt{\varepsilon_{4+1} \mu_{4+1}}} = \frac{1}{\sqrt{\varepsilon_0 \mu_0}} \tag{73}$$

In $(N+1)$-D, Coulomb's law is a $1/r^{N-1}$ force law if the extra dimensions are noncompact. If instead there is toroidal compactification and the extra dimensions are symmetric – i.e. they all have the same radius $R$ – then, in the limit that $R_3 \gg R$, Coulomb's law is approximately a $1/r^2$ force law in the effective $(3+1)$-D theory. In this case, the $(N+1)$-D permittivity of free space $\varepsilon_{N+1}$ is related to the usual $(3+1)$-D permittivity $\varepsilon_0$ via

$$\varepsilon_{N+1} = \frac{\varepsilon_0}{\left(2R\sqrt{\pi}\right)^{N-3}\Gamma\left(\frac{N-3}{2}\right)} \qquad (74)$$

where $N > 3$.

The deviation in the usual $1/r^2$ form of Coulomb's law can be computed for a specified geometry of extra dimensions through the lowest-order corrections to the integration in Eq. 70 [11]. The effects are similar to the deviation in the usual $1/r^2$ form of Newton's law of universal gravitation.

## 8. Conclusions

We considered in detail how the standard physics formulas involving the curl operation and cross product can be physically viable in higher-dimensional theories when it is well-known that the cross product is only valid in $(3+1)$-D and $(7+1)$-D, but physically meaningful only in $(3+1)$-D. We assert that, in all of the standard physics formulas involving the curl operation and cross product, one of the quantities normally identified as a vector is actually an anti-symmetric second-rank tensor. We presented two alternative tensor constructions (Eq.'s 2 and 4), which are equivalent to the usual cross product in $(3+1)$-D, and illustrated how to reformulate the standard theories in terms of these anti-symmetric second-rank tensors. Unlike the conventional cross product between vectors, the tensor constructions naturally generalize to higher dimensions, which we illustrated with examples.

We also considered how the differential elements are impacted in Gauss's law and Stokes's theorem in higher dimensions. We found that the most natural forms of the various differential elements are: a vector $d\vec{\mathbf{s}}$ for the line element, an anti-symmetric second-rank tensor $d\underline{\underline{\mathbf{A}}}$ for the differential area, a third-rank tensor $d\underline{\underline{\mathbf{V}}}$ for the differential volume, which can also be expressed in terms of a vector $d\vec{\mathbf{V}}$ in $(4+1)$-D, and so on. The higher-dimensional analog of the differential path length can be expressed as an anti-symmetric second-rank tensor, while the higher-dimensional analog of the surface area can be expressed as a vector.

Alternatively, Gauss's law and Stokes's theorem may be generalized in the conventional way using the differential forms. On the one hand, this is a very powerful geometric formalism, where the coordinates are completely suppressed, which is very useful for problems with a high degree of symmetry. Our tensor construction provides some advantages when Gauss's law and Stokes's theorem are expressed in terms of physical fields in Maxwell's equations. We provide direct integrals for the vector electric field $\vec{\mathbf{E}}$ and tensor magnetic field $\underline{\underline{\mathbf{B}}}$, clearly showing the nature of these fields, that is practical for problems without much symmetry. Our form also closely parallels the usual vector calculus construction of Maxwell's equations.

We applied the tensor constructions of the usual cross products and higher-dimensional generalizations of the differential elements of vector calculus specifically to generalize Maxwell's equations in differential and integral form to higher dimensions.

We found that the $(N+1)$-D field strength tensor naturally divides into an $N$-component vector electric field $\vec{\mathbf{E}}$ and an $N \times N$ anti-symmetric second-rank tensor magnetic field $\underline{\mathbf{B}}$ with $N(N-1)/2$ components. The usual curl of the electric field $\vec{\nabla} \times \vec{\mathbf{E}}$ also generalizes to an anti-symmetric second rank tensor $\partial_I E_J - \partial_J E_I$. We then applied Maxwell's equations in higher dimensions to a couple of classic $(3+1)$-D examples. We found that the usual inverse-square laws in $(3+1)$-D become $1/r^{N-1}$ laws in $(N+1)$-D. We also saw that the case for magnetic monopoles is not as aesthetically compelling in higher dimensions as it is in $(3+1)$-D because the usual correspondence between the electric and magnetic fields – i.e. $\vec{\mathbf{E}}/c \to \vec{\mathbf{B}}$ and $\vec{\mathbf{B}} \to \vec{\mathbf{E}}/c$ – is coincidental in $(3+1)$-D.

Beginning with the integral form of Maxwell's equations in $(4+1)$-D, we compactified the extra dimension $x^4$ on an $S^1/Z_2$ orbifold with the orbifold symmetry $x^4 \to -x^4$. We obtained the effective $(3+1)$-D theory, which includes the usual form of Maxwell's equations plus additional terms involving KK excitations, by first Fourier-expanding the $(4+1)$-D fields and then integrating over the compactified dimension $x^4$. The current collider bound on the compactification scale $1/R$ indicates that $1/R$ must exceed ~350-400 GeV for universal extra dimensions. We also illustrated how the universal extra dimensions impact the form of Coulomb's inverse-square law, and related the usual $(3+1)$-D permittivity and permeability of free space, $\varepsilon_0$ and $\mu_0$, respectively, to their higher-dimensional counterparts, $\varepsilon_{N+1}$ and $\mu_{N+1}$.

AM would like to thank the Louisiana School for Math, Science, and the Arts for their kind hospitality during this research.